\documentclass[review]{elsarticle}
\graphicspath{ {./figures/} }
\usepackage{hyperref}
\usepackage{float}
\usepackage{verbatim} 
\usepackage{apalike}
\restylefloat{figure}
\restylefloat{table}

\usepackage{graphicx,subcaption}
\usepackage{amsmath}
\usepackage{amssymb}

\usepackage{lineno}

\usepackage{xcolor}
\usepackage{colortbl}
\usepackage{array}
\usepackage{pifont}
\usepackage{multirow}
\usepackage{hyperref}
\usepackage{framed}

\usepackage{tabularx}
\usepackage{enumitem}
\setlist{nolistsep}
\definecolor{green}{HTML}{66FF66}
\definecolor{novabarva}{HTML}{009700}

\journal{Journal}

\biboptions{authoryear}

\begin{document}
\begin{frontmatter}

\title{A comprehensive review of visualization methods for association rule mining: Taxonomy, Challenges, Open problems and Future ideas}

\author[label1,label2]{Iztok Fister Jr.}
\ead{iztok.fister1@um.si}

\author[label1]{Iztok Fister}
\ead{iztok.fister@um.si}

\author[label1]{Du\v{s}an Fister \corref{cor1}}
\ead{dusan.fister@um.si}

\author[label1]{Vili Podgorelec}
\ead{vili.podgorelec@um.si}

\author[label2]{Sancho Salcedo-Sanz}
\ead{sancho.salcedo@uah.es}

\cortext[cor1]{Corresponding author.}
\address[label1]{Faculty of Electrical Engineering and Computer Science, University of Maribor, Koro\v{s}ka cesta 46, 2000 Maribor, Slovenia.}
\address[label2]{Department of Signal Processing and Communications, Universidad de Alcal\'a,  Alcal\'a de Henares, Madrid, 28805, Spain}

\begin{abstract}
Association rule mining is intended for searching for the relationships between attributes in transaction databases. The whole process of rule discovery is very complex, and involves pre-processing techniques, a rule mining step, and post-processing, in which visualization is carried out. Visualization of discovered association rules is an essential step within the whole association rule mining pipeline, to enhance the understanding of users on the results of rule mining. Several association rule mining and visualization methods have been developed during the past decades. This review paper aims to create a literature review, identify the main techniques published in peer-reviewed literature, examine each method's main features, and present the main applications in the field. Defining the future steps of this research area is another goal of this review paper.
\end{abstract}


\end{frontmatter}

\section{Introduction}
\label{sec:intro}
Association Rule Mining (ARM) is definitely one of the most important and popular data mining techniques for discovering unknown knowledge from transaction databases. The ARM is also a part of Machine Learning (ML) with the task to discover interesting relationships between items in large transaction datasets. The relationships are expressed by association rules determining how and why certain items are connected. The story of ARM started with a seminal paper of~\cite{agrawal1994fast}. Agrawal set the theoretical foundations for the process of ARM, and proposed the first algorithm, called {{\em Apriori}. Apriori is a deterministic algorithm for mining association rules, and is still today featured as one of the top algorithms in the DM domain~\citep{wu2008top}, as well as a member of an unprecedented scale in student textbooks.  

In the following years, the ARM gained huge interest in the ML community. Its popularity was proven with many practical applications, especially, in the domains such as market-based analysis~\citep{nisbet2018advanced}, medical diagnosis~\citep{xu2022privacy}, census data~\citep{malerba2002mining} or protein sequences~\citep{gupta2006mining}, among others.

Data analysis pipelines typically consist of data cleanup and minimizing data imputations (also data pre-processing), data collection and exploration design, and comprehending the mined knowledge. Thus, the whole ARM pipeline is complex (see Fig.~\ref{basic-pipeline}), because it consists of three steps, as follows: the pre-processing, the ARM, and the post-processing. The input to the pipeline presents the transaction database, which consists of rows and columns, where each row presents a transaction, and columns the attributes. In the pre-processing step, some optional substeps can be applied to make the data more robust, i.e., data cleaning and missing data imputation, where some outliers, or rows with a lot of missing data can even be removed. On the other hand, some other operations, for example, data squashing~\citep{fister2022datasqa}, can help reduce the transaction dataset. Then, the ARM process itself is performed. In line with this, several algorithms, e.g., Apriori or Eclat, exist and are, as mentioned, some of the most used. The output of this step is usually a huge collection of mined/identified/found association rules. Usually, researchers present these rules as a table, or summarize them using some metrics. However, visualization of the association rules needs to be conducted for the best insights. 

\begin{figure}
\centering
\includegraphics[width=9cm]{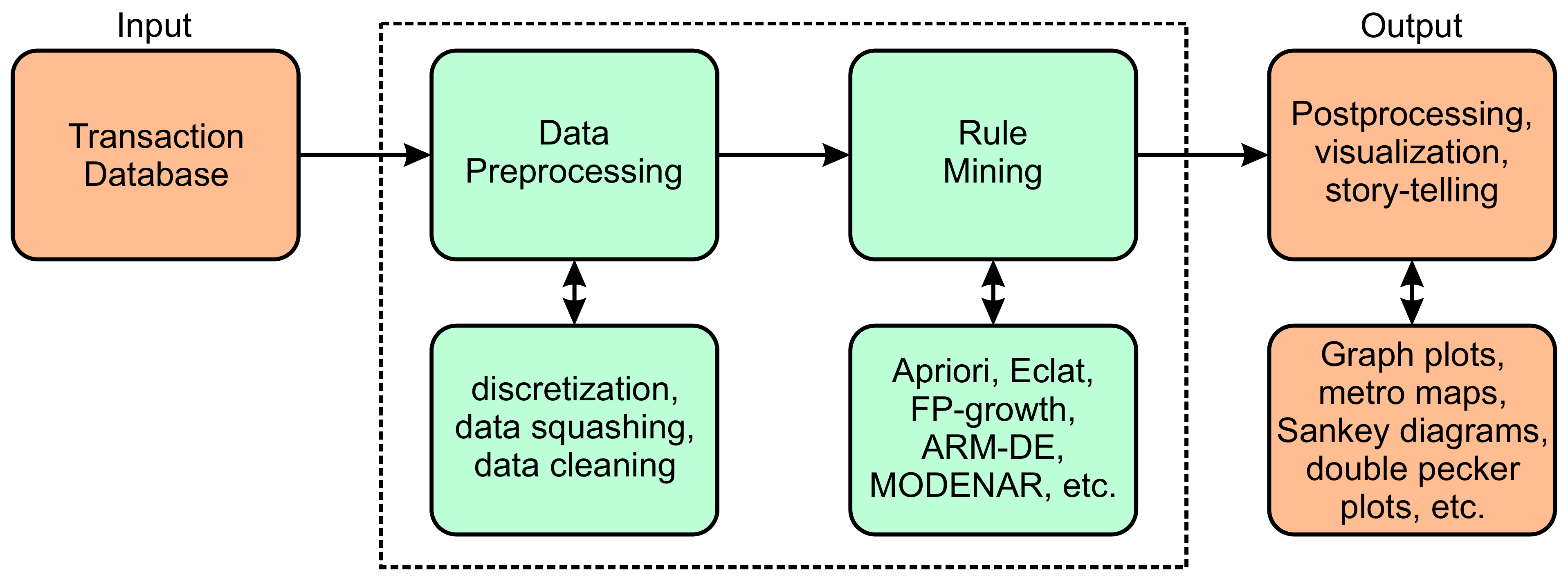}
\caption{The basic ARM pipeline.}
\label{basic-pipeline}
\end{figure}

Nowadays in ML, there is a trend to go for easier representation of the results obtained by ML/AutoML (automated ML) pipelines. This intention also coincides with the emerging research area of eXplainable Artificial Intelligence (XAI)~\citep{barredo2019explainable,arrieta2020explainable}. XAI has become an important part of the future of AI, because XAI models explain the reasoning behind their decisions. This provides an increased level of understanding between humans and machines, which can help build trust in AI systems~\citep{kumar2022what}. In summary, XAI is a set of processes and methods to comprehend and trust the results created by the ML algorithms. In line with this, it tries to describe an AI model's impact on the one hand, and exposes its potential biases on the other. Thus, the ML model is estimated according to its accuracy, fairness, transparency, and outcomes of AI-powered decision-making~\citep{borego2022explainable}. 

XAI can be manifested in several forms: text explanation, visualization, local explanation, explanation by example, explanation by simplification, and feature relevance~\citep{barredo2019explainable,bennetot4229624practical}. Thus, there is an increased interest of researchers in developing new methods for easier representation of the results. Definitely, one of the most important parts of these efforts is visualization methods \citep{arrieta2020explainable}.  

Typically, ARM algorithms generate a huge number of association rules. Frequently, the results are opaque for ordinary users, and need some explanations to understand their meaning. On the other hand, visualization of the results has a huge explanation power. Although a lot of visual methods have been proposed for ARM, to the best of our knowledge, no review for dealing with this problem from the XAI point of view exists nowadays. 

Therefore, the aim of this paper is to collect and discuss visualization techniques for ARM that have appeared from its advent to the present day. Each method is studied in detail and features are compared with each other in the sense of XAI. The contributions of this review paper are summarized as follows:
\begin{itemize}
    \item The evolution of ARM visualization methods is presented.
    \item The features of each of the methods are defined.
    \item The advantages/disadvantages of each method are outlined.
    \item An example is presented for each of the surveyed methods.
    \item Explaining models using the ARM visualization are summarized.
\end{itemize}

The review of the ARM visualization methods is based on papers published from three different main sources: the ACM Digital Library, IEEEXplore, and Google Scholar. The analyse of the methods are highlighted from the following points of view: (1) characteristics, (2) visualization focus, and (3) attribute type. The taxonomies of the ARM visualization methods are introduced based on the highlights. 

The structure of the paper is organized as follows: Section~\ref{sec:2} deals with the ARM problem in a nutshell. The mathematical definition of the ARM visualization is the subject of Section~\ref{sec:3}. A detailed overview of traditional ARM visualization methods is reviewed in Section~\ref{sec:4}. New ideas in the ARM visualization are the subject of Section~\ref{sec:5}. The subject of Section~\ref{sec:6} is a review of graphical systems, while Section~\ref{sec:7} introduces challenges and open problems. The review concludes with Section~\ref{sec:8} that summarizes the performed work and outlines potential ideas for the future work.

\section{Association rule mining in a nutshell}\label{sec:2}
The ARM problem is defined formally as follows: Let us suppose a set of items $I=\{i_1, \ldots,i_M\}$ and transaction database $D=\{Tr_1,\ldots,Tr_N\}$ are given, where each transaction $Tr_i$ is a subset of objects $Tr_i \subseteq I$. Thus, the variable $M$ designates the number of items, and $N$ the number of transactions in the database. Then, an association rule can be defined as an implication:
\begin{equation}
X \Rightarrow Y, 
\label{arule}
\end{equation}
\noindent where $X \subset I$  (left-hand-side or LHS), $Y \subset I$  (right-hand-side or RHS), and $X \cap Y = \emptyset$. The following four measures are defined for evaluating the quality of the association rule~\citep{agrawal1994fast}:
\begin{equation}
\mathit{supp}(X \Rightarrow Y) = \frac{n(X \cap Y)}{N}, 
\label{Eq:2}
\end{equation}
\begin{equation}
\mathit{conf}(X \Rightarrow Y) = \frac{n(X \cap Y)}{n(X)}, 
\label{Eq:1}
\end{equation}
\begin{equation}
\mathit{lift}(X \Rightarrow Y) = \frac{supp(X \cap Y)}{supp(X)\times supp(Y)}, 
\label{Eq:lift}
\end{equation}
\begin{equation}
\mathit{conv}(X \Rightarrow Y) = \frac{1-supp(Y)}{1-conf(X\Rightarrow Y)}, 
\label{Eq:conv}
\end{equation}
\noindent where $\mathit{supp}(X \Rightarrow Y) \geq S_{\mathit{min}}$ denotes the support, $\mathit{conf}(X \Rightarrow Y)\geq C_{\mathit{min}}$ the confidence, $\mathit{lift}(X \Rightarrow Y)$ the lift, and $\mathit{conv}(X \Rightarrow Y)$ the conviction of the association rule $X \Rightarrow Y$. There, $N$ in Eq.~(\ref{Eq:2}) represents the number of transactions in the transaction database $D$, and $n(.)$ is the number of repetitions of the particular rule $X \Rightarrow Y$ within $D$. Additionally, $S_{\mathit{min}}$ denotes minimum support and $C_{\mathit{min}}$ minimum confidence, determining that only those association rules with confidence and support higher than $C_{\mathit{min}}$ and $S_{\mathit{min}}$ are taken into consideration, respectively. 

The interpretations of the measures are as follows: The support measures the proportion of transactions in the database which contain the item. The confidence estimates the conditional probability $P(Y|X)$, denoting the probability to find the $Y$ of the rule in transaction under the condition that this transaction also contains the $X$. The lift is the ratio of the observed support that $X$ and $Y$ arose together in the transaction if both set of items are independent. The conviction evaluates the frequency with which the rule makes an incorrect prediction.

\subsection{Numerical association rule mining}
Numerical Association Rule Mining (NARM) extends the idea of ARM, and is intended for mining association rules where attributes in a transaction database are represented by numerical values. Usually, traditional algorithms, e.g., Apriori, require a discretization of numerical attributes before they are ready to use. The discretization is sometimes trivial, and thus does not affect the results of mining positively. 

On the other hand,  many methods for ARM exist that do not require the discretization step before applying the process of mining. Most of these methods are based on population-based nature-inspired metaheuristics, such as, for example, Differential Evolution (DE)~\citep{storn1997differential} or Particle Swarm Optimization (PSO)~\citep{kennedy1995particle}.  Consequently, the NARM has recently showed an importance in the data revolution era that has been confirmed by some review papers~\citep{altay2019performance,telikani2020survey} tackling the solving this class of problems.

Each numerical attribute is determined in NARM by an interval of feasible values limited by its lower and upper bounds. The more association rules are mined the broader the interval. The narrower the interval, the more specific relations are discovered between attributes. Introducing intervals of feasible values has two major effects on the optimization: To change the existing discrete search space to continuous, and to adapt these continuous intervals to suit the problem of interest better.

Mined association rules can be evaluated according to several criteria, like support and confidence. For the NARM, however, additional measures must be considered, in order to evaluate the mined set of association rules properly.

\subsection{Time Series Association Rule Mining}
TS-ARM is a new paradigm, which treats a transaction database as a time series data. The formal definition of the NARM problem needs to be redefined in line with this. In the TS-ARM, the association rule is defined as an implication:
\begin{equation}
    X(\Delta t)\implies Y(\Delta t),
\end{equation}
where $X(\Delta t)\subset O$, $Y(\Delta t)\subset O$, and $X(\Delta t)\cap Y(\Delta t)=\emptyset$. The variable $\Delta t=[t_1,t_2]$ determines the sequence of the transactions which have arisen within the interval $t_1$ and $t_2$, where $t_1$ denotes the start and $t_2$ the end time of the observation. The measures of support and confidence are redefined as follows:
\begin{equation}
\mathit{conf_t}(X(\Delta t) \implies Y(\Delta t)) = \frac{n(X(\Delta t) \cap Y(\Delta t))}{n(X(\Delta t))}, 
\label{Eq:1t}
\end{equation}
\begin{equation}
\mathit{supp_t}(X(\Delta t) \implies Y(\Delta t)) = \frac{n(X(\Delta t) \cap Y(\Delta t))}{N(\Delta t)}, 
\label{Eq:2t}
\end{equation}
where $\mathit{conf_t}(X(\Delta t) \implies Y(\Delta t))\geq C_{\max}$ and $\mathit{supp_t}(X(\Delta t) \implies Y(\Delta t))\geq S_{\max}$ denotes the confidence and support of the association rule $X(\Delta t)\implies Y(\Delta t)$ within the same time interval $\Delta t$. 

\section{Visualization of association rule mining}\label{sec:3}
Visualization of ARM can be described mathematically as a set of triplets:
\begin{equation}
    \mathcal{R}=\{\langle X_1,Y_1,Z_1\rangle,\ldots,\langle X_i,Y_i,Z_i\rangle,\ldots,\langle X_n,Y_n,Z_n\rangle \},
\end{equation}
where $X_i$ denotes an antecedent, $Y_i$ a consequent, and $Z_i$ a vector of available interestingness measures (e.g., support, confidence, etc.) for $i=1,\ldots,N$. In a nutshell, different visualization methods depend on: 
\begin{itemize}
\item the number of interstingness measures to display,
\item the visualization focus,
\item the rule set size.
\end{itemize}
The number of interestingness measures to display is limited by the number of dimensions that can be visualized (i.e., 2D or 3D). The visualization focus determines how the association rule defines the neighborhood of rules to be visualized. In line with this, the neighborhood is defined by: interestingness measure, items, similarity of RHS and LHS, or time series' visualization. The rule set size limits the number of association rules that are included into a specific visualization method.

\subsection{Study design}
\label{res-methodology}
For conducting the systematic literature review, we followed the guidelines presented in the Systematic Literature Review Guidelines in Software Engineering \citep{kitchenham2007guidelines}. Our primary goal was to identify the frequency of the ARM visualization methods, the main features of these methods, and the applications in which these methods were applied. According to our goals, we developed the following Research Questions (RQ)s:
\begin{itemize}
\item RQ1: Which methods are developed for the ARM visualization?
\item RQ2: Which challenges and open problems are placed behind the ARM visualization? 
\item RQ3: Which software packages are available to users tackling these problems?
\item RQ4: What awaits the methods for visualization of association rules in the future? 
\end{itemize}
We conducted a literature search using major databases from 18 to 22 November, 2022. The main search strings that were used for searching the databases were as follows: ``association rule mining'' AND ``visualization'' OR ``visualisation''. The search string was also modified according to the search formats of different databases.

Table~\ref{tab:papers} presents the results of our search~\footnote{Note that we also checked the citing articles of results from Google Scholar manually.}. Each of the papers was prescreened according to its abstract and keywords. 
\begin{table}[htb]
  \caption{Search results of papers regarding the keywords in various databases.}
  \label{tab:papers}
  \centering
  \begin{tabular}{ l l r r }
\hline
 \textbf{Database name} & \textbf{URL} & \textbf{Total} & \textbf{Included} \\ 
 \hline
 ACM Digital Library & \href{https://dl.acm.org}{dl.acm.org} & 6 & 4 \\ 
 IEEEXplore & \href{https://ieeexplore.ieee.org}{ieeexplore.ieee.org} & 214 & 21 \\
 Google Scholar & \href{https://scholar.google.com}{scholar.google.com} & 16,100  & 25+ \\ 
\hline
 Total & & 16,320 & 25+ \\
 \hline
\end{tabular}
\end{table}

When the results were collected, we also filtered out the duplicates. Additionally, when searching through the Google scholar we checked for citing articles of each paper, so that additional results were then identified and included in this review paper. We also specified the selection and exclusion criteria as well as limitations. The selection criteria were the follows: (1) research paper addresses any kind of ARM and its connection with visualization, and the research must be peer reviewed, i.e., published in a referred conference, journal paper, book chapter or monograph. The search was conducted with exclusion criteria as follows: ``The research paper is not written in the English language'', and limitations such as: ``The literature review search was limited to only three databases''. The summary of abstracts from IEEEXplore and ACM Digital Library publications is shown in the wordcloud~Fig.~\ref{wordcloud}.

\begin{figure}
\centering
\includegraphics[width=12cm]{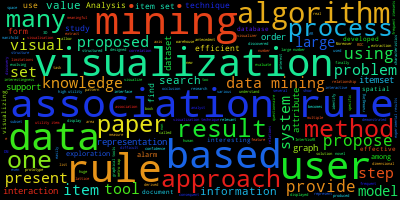}
\caption{Wordcloud of the extracted abstracts.}
\label{wordcloud}
\end{figure}

\section{A detailed overview of traditional ARM visualization methods}\label{sec:4}
In the following subsections, each of the methods is outlined, followed by a summary of related work, while several methods are also illustrated by an example. The examples of the particular visualization are implemented in arulesViz \citep{hahsler2011arules} on a set of 11,267 association rules produced by the Apriori algorithm \citep{agrawal1994fast} mining the Mushroom UCI ML dataset \citep{uci2020repository} using the following limitations: $S_{\mathit{min}}=0.3$ and $S_{\mathit{min}}=0.5$. 

Table~\ref{tab:1} presents a summary of the traditional ARM visualization methods that were found in our systematic literature review. It is divided into four columns that present: a sequence number (column 'Nr.'), a class (column 'Class'), a variant (column 'Variant'), and the method's developer (column 'Reference').
\begin{table}[htb]
\caption{Summary of ARM visualization methods.}
\label{tab:1}
\begin{tabular}{ c|l|l|l }
\hline
Nr. & Class & Variant & Reference \\\hline
 \multirow{2}{*}{1} & \multirow{2}{*}{Scatter} & Scatter plot & \citet{bayardo1999mining} \\ 
 &  & Two key plot & \citet{unwin2001twokey} \\ \hline 
 \multirow{1}{*}{2} & \multirow{1}{*}{Graph} & Graph-based & \citet{klemettinen1994finding} \\\hline 
 \multirow{2}{*}{3} & \multirow{2}{*}{Matrix} & Matrix-based & \citet{Ong2002crystalClear} \\
  & & Grouped matrix-based & \citet{hasler2017visualizing} \\\hline 
 \multirow{2}{*}{4} & \multirow{2}{*}{Mosaic} & Mosaic plot & \citet{Hofmann2008mosaic} \\ 
  &  & Double decker plot & \citet{hofman2001visual} \\\hline  
 \hline
\end{tabular}
\end{table}
As can be seen from the table, we are focused on eight classes of visualization methods and their variants (together 7 visualization methods). In the remainder of the paper, the aforementioned visualization methods are illustrated in a nutshell.

\subsection{Scatter plot}
A Scatter plot (Fig.~\ref{fig:Scatter}) was firstly used for visualizing mined association rules by \citep{bayardo1999mining}. 
\begin{figure}[htb]
\centering
  \begin{subfigure}[t]{.48\linewidth}
    \includegraphics[width=1.\textwidth]{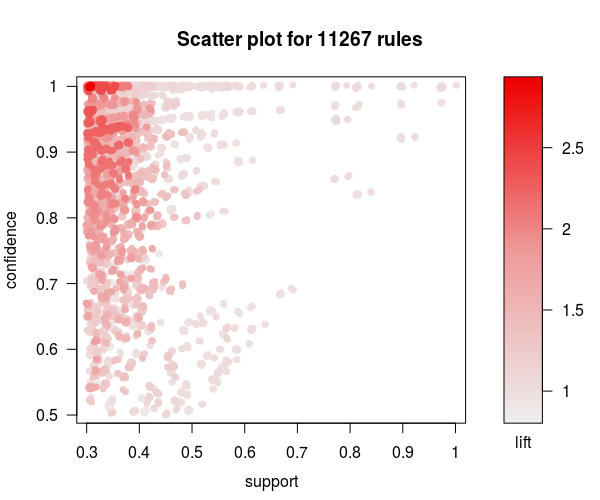}
    \caption{Scatter plot.}
    \label{fig:Scatter}
  \end{subfigure}
  \begin{subfigure}[t]{.48\linewidth}
  \centering
    \includegraphics[width=1.\linewidth]{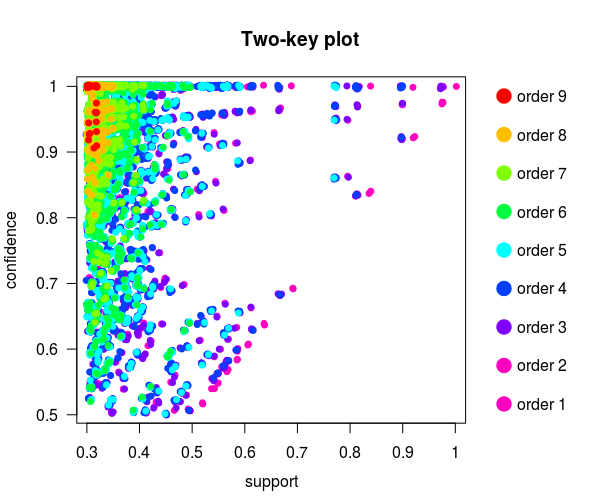}
    \caption{Two-key plot.}
    \label{fig:Twokey}
  \end{subfigure}
  \caption{Scatter and Two-key plots powered by arulesViz.}
  \label{fig:Scatter_Two-key}
\end{figure}
In general, this plot is used to display an association or relationship between interestingness measures $Z_i$ (usual support and confidence) that are presented as dots in the Scatter plot. Additionally, the third measure (usual lift) is included as a color key. Thus, rules with similar values of interestingness measures are placed closer to each other, while the correlation can be established between dependent and independent variables. Typically, the so-called regression line is drawn in the Scatter plot, representing the trend of the relationship between two observed variables. This line can also be used as a predictive tool in some circumstances.

\subsubsection{Twokey plot}
A two-key plot (Fig.~\ref{fig:Twokey}) is a special kind of Scatter plot that was developed by \cite{unwin2001twokey}, especially, for analyzing association rules. It consists of a two dimensional Scatter plot displaying an association between two measures of interestingness (usually support and confidence), while the third measure is represented by the color of the points (i.e., support/confidence pairs), where the color corresponds to the length of the rule (also order). Interestingly, 2-order association rules describe trails moving from the upper right side (perfect result) to the left lower side of the same plot (lesser support and lesser confidence).

\subsection{Graph-based}
Graph-based techniques (Fig.~\ref{fig:graph}) identify how rules share individual item \citep{klemettinen1994finding,rainsford2000temporal,buono2005visualizing,ertek2006framework}. They visualize \begin{figure}[htb]
\centering
\begin{minipage}{.5\textwidth}
  \centering
  \includegraphics[width=1.\textwidth]{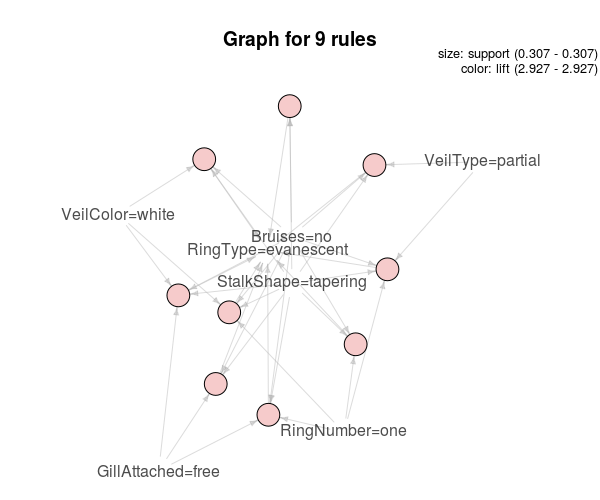}
  \caption{Graph plot powered by arulesViz.}
  \label{fig:graph}
\end{minipage}%
\end{figure}
association rules using vertices and edges, where vertices annotated with item labels represent items, and itemsets or rules are represented as a second set of vertices. The items are connected with itemsets/rules using arrows. For rules, arrows pointing from items to rule vertices indicate LHS items, and an arrow from a rule to an item indicates the RHS. Interestingness measures are typically added to the plot by using the color or the size of the vertices representing the itemsets/rules. Graph-based visualization offers a very clear representation of rules but they tend to become cluttered easily, and, thus, are only viable for very small sets of rules. 

\subsection{Matrix-based}
Matrix-based visualization \citep{Ong2002crystalClear} (Fig.~\ref{fig:Matrix}) identifies associations between antecedent (LHS) and consequent (RHS) items. 
\begin{figure}[htb]
\centering
\begin{subfigure}[t]{.48\linewidth}
  \centering
  \includegraphics[width=1.\linewidth]{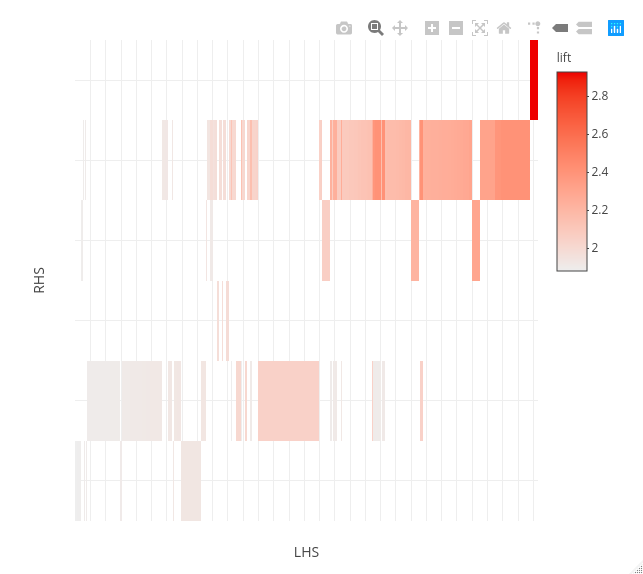}
  \caption{Matrix plot.}
  \label{fig:Matrix}
\end{subfigure}
\begin{subfigure}[t]{.48\linewidth}
  \centering
  \includegraphics[width=1.\textwidth]{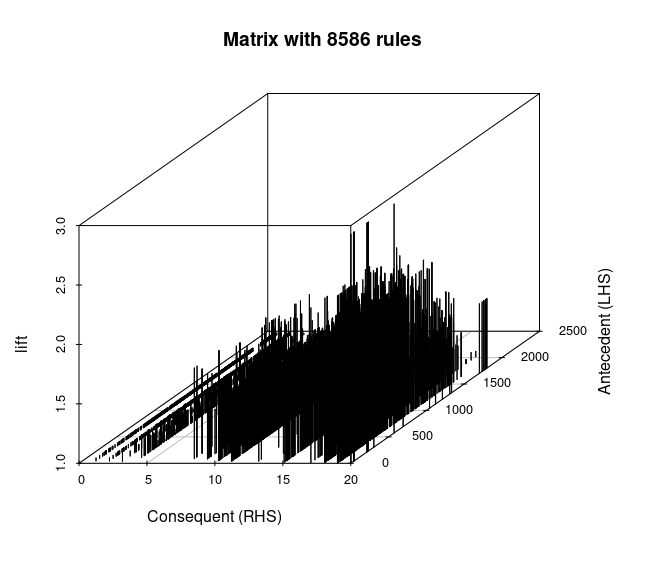}
  \caption{Grouped matrix-based plot.}
  \label{fig:Matrix3D}
\end{subfigure}
\caption{Matrix and Grouped matrix-based plots powered by arulesViz.}
\label{fig:Matrix2}
\end{figure}
Thus, association rules are organized as a square matrix $M=\{m_{j,k}\}$ of dimension $M\times M$, in which distinct antecedent items $X_i\in\{x_{i,j}\}$ for $j=1,\ldots,|X_i|$ and distinct consequent items $Y_i\in\{y_{i,k}\}$ for $k=1,\ldots,|Y_i|$ are included. The values of some interestingness measure (e.g., lift) are then assigned to the corresponding position $m_{j,k}=Z_i$  of the matrix. Typically, the antecedent itemset of the rules is ordered by increasing support, while the consequent itemset by increasing confidence before visualization.

However, the matrix visualization is limited by the rule set size (i.e., $<1,000$), especially in the case of a huge matrix, which makes the exploration of the matrix much harder.

\subsubsection{Grouped matrix-based visualization}
The grouped matrix-based visualization \citep{hasler2017visualizing} (Fig.~\ref{fig:Matrix3D}) is a variant of the original matrix-based visualization, where the large set of different antecedents (the columns in matrix $M$) are grouped into the smaller set of groups using clustering. Mathematically, the set of antecedents is grouped into a set of $k$ groups $S=\{S_1,\ldots,S_k\}$ according to minimizing the sum of squares within the particular cluster, in other words:
\begin{equation}
\arg\min_S\sum_{i=1}^k\sum_{m_{i,j}\in S_i} ||m_{i,j}-\overline{m}_i||^2,
\end{equation}
where $\mathbf{m}_i=\{m_{i,j}\}$ for $j=1,\ldots,|A_i|$ is a column $i$ of matrix $\mathbf{M}$ which represents all values with the same antecedent, and $\overline{m}_i$ is the center of the cluster $S_i$. Thus, the $k$-means algorithm~\citep{hartigan1979k-means} is applied 10-times with random initialization of the centroids. The best solution is then used for an ARM visualization. The motivation behind the ARM visualization method is to reduce the antecedent's dimension that enables more informative visualization of the association rules.

\subsection{Mosaic plot}
A mosaic plot \citep{hartigan1984mosaic} is applied for visualizing the interesting rule, consisting primarily of categorical attributes (Fig.~\ref{fig:Mosaic}). It is based on the so-called contingency table, in which the frequencies of the attribute appearances in the interesting rule $r^*$ are assigned to each position $m_{j,k}$, where $j$ denotes the corresponding the antecedent attribute $A_j$ and $k$ the consequent attribute $A_k$. 

\begin{figure}[htb]
\centering
\begin{subfigure}[t]{.48\linewidth}
  \centering
  \includegraphics[width=1.\linewidth]{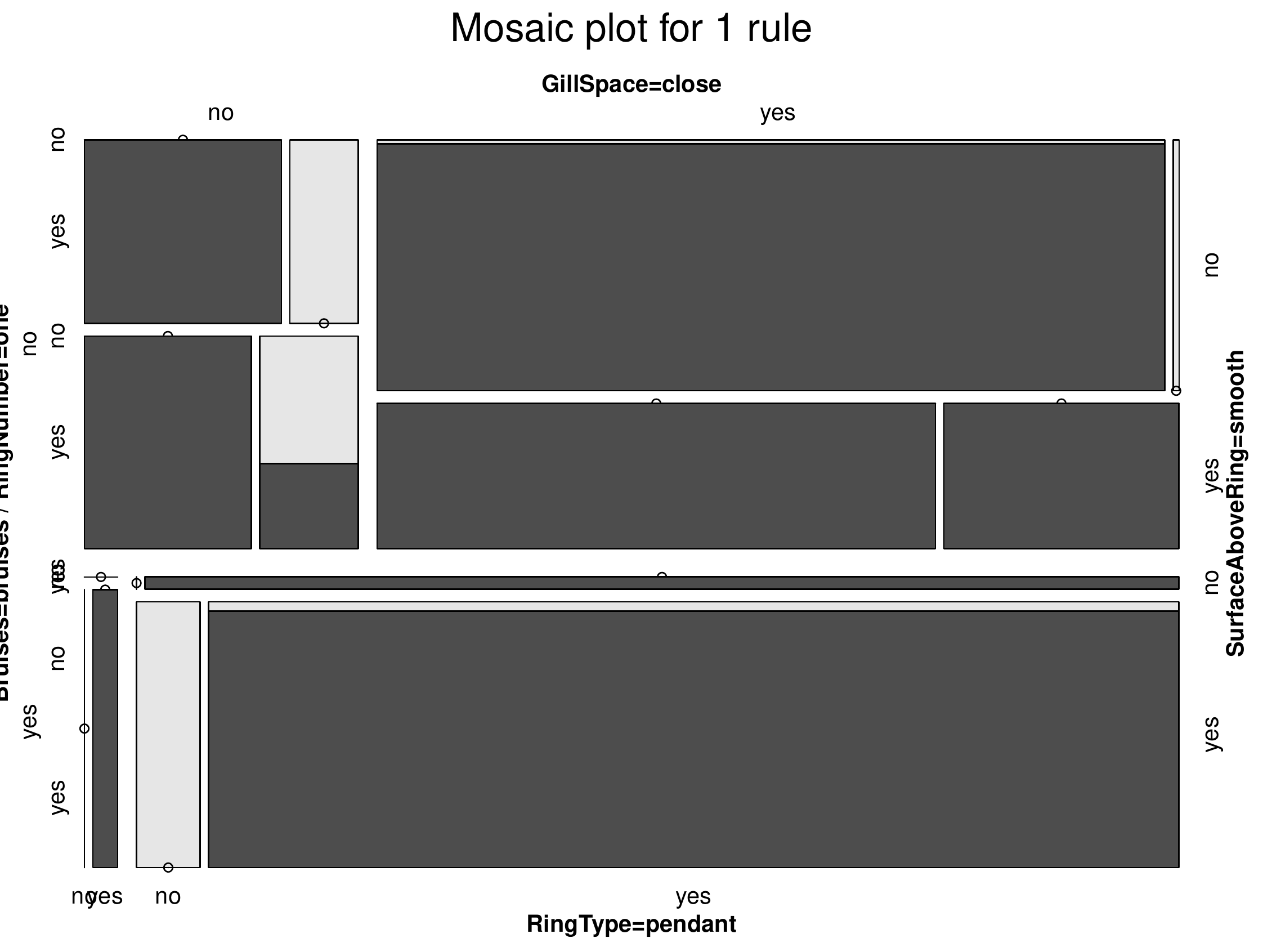}
  \caption{Mosaic plot.}
  \label{fig:Mosaic}
\end{subfigure}
\begin{subfigure}[t]{.48\linewidth}
  \centering
  \includegraphics[width=1.\textwidth]{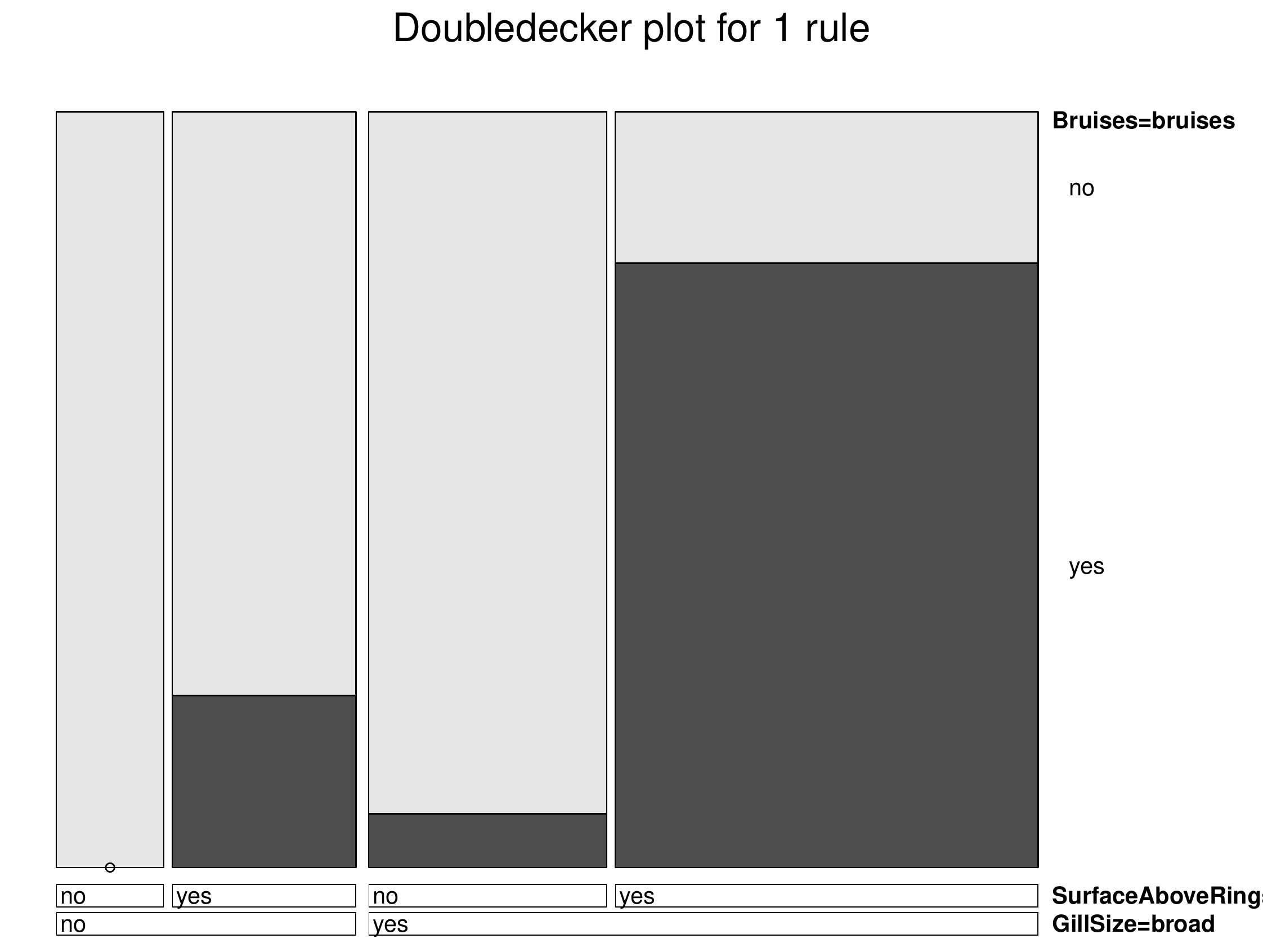}
  \caption{Double Decker plot.}
  \label{fig:DoubleDecker}
\end{subfigure}
\caption{Mosaic and Double Decker plots powered by arulesViz.}
\label{fig:DoubleDecker2}
\end{figure}

The interesting rule is determined as follows: Let us assume that each rule $r_i\in \text{R}$ is a tuple $r_i=\langle X_i,Y_i,Z_i\rangle$, where $X$ denotes the attributes $A=\{A_i,\ldots,A_p\}$ belonging to the antecedent, $Y$ to the consequent, $Z$ is a set of interestingness measures, and $X\cap Y=\emptyset$. Then, the interesting rule $r^*$ for visualizing with mosaic plot is defined as
\begin{equation}
    r^*\Rightarrow Y|Z,
\end{equation}
where $X=\{A_{x_1}=a_{x_1}\wedge A_{x_k}=a_{x_k}\}$, $Y=\{A_{y}=a_{y}\}$, and $Z=\{\mathit{supp},\mathit{conf}\}$, for which the difference of confidence (doc) for rule $X\Rightarrow Y$ and $\neg X\Rightarrow Y$ is the maximum, in other words:
\begin{equation}
    \max_{r\in\text{R}}\mathit{conf}(X\Rightarrow Y)-\mathit{conf}(\neg X\Rightarrow Y).
\end{equation}
Mosaic plots were introduced as a graphical analogy of multivariate contingency tables \citep{hofman2000visualizing}. This means that the position $m_{i,j}$ (also a cell in a contingency table) is presented in a mosaic plot as an area divided into the highlighted part (colored) that is proportional to the support of the rule $X\Rightarrow Y$ and the unhighlighted part of the rule $\neg X\Rightarrow Y$. Thus, the confidence is proportional to the height of the highlighted part of the area.

\subsubsection{Double Decker plot}
Double Decker plot \citep{hofmann2000mosaic} allows comparing the proportions of the highlighted heights referring to confidence measure more easily (Fig.~\ref{fig:DoubleDecker}). While the original mosaic plot splits tiles in vertical and horizontal directions, the Double Decker splits these only horizontally. As a result, the antecedent of the interesting rule is now expressed mathematically as: 
\begin{equation*}
X=\{A_{x_1}=\cdot\wedge A_{x_p}=\cdot\},    
\end{equation*}
i.e., the proportions of the highlighted heights are presented in each tile of the mosaic plot, while the widths of the tiles are represented as labels denoting the antecedent's attributes. Thus, the highlighted shades illustrate relations with an outcome set to 'True', while the white shades refer to relations, whose outcome leads to 'False'. 

\section{New ideas in the visualization of association rules}\label{sec:5}
This section reviews papers dealing with ARM visualization methods that accumulate new ideas in this domain. The ideas are collected in Table~\ref{tab:1a},
\begin{table}[htb]
\caption{Summary of the new ARM visualization methods.}
\label{tab:1a}
\begin{tabular}{ c|l|l|l }
\hline
Nr. & Class & Variant & Reference \\\hline
 1 & \multirow{1}{*}{Fishbone} & Ishikawa diagram & \citet{tsurinov2021farm} \\\hline 
 2 & \multirow{1}{*}{Molecular} & Molecular representation & \citet{said2013visualisation} \\\hline 
 3 & \multirow{1}{*}{Lattice} & Concept lattice & \citet{shen2020research} \\\hline 
 4 & \multirow{1}{*}{Metro} & Metro map & \citet{fister2022information} \\\hline 
 5 & \multirow{1}{*}{Sankey} & Sankey diagram & \citet{fister2022association} \\\hline 
 6 & \multirow{1}{*}{Ribbon} & Ribbon plot & \citet{fister2020visualization} \\\hline 
 7 & \multirow{1}{*}{Glyph} & Glyph-based & \citet{hrovat2015interestingness} \\\hline 
 \hline
\end{tabular}
\end{table}
from which it can be seen that, here, we were focused on the seven ARM visualization methods, which, in our opinion, best reflect the development in this domain. In the remainder of the section, the selected ARM visualization methods are illustrated in detail.

\subsection{Ishikawa diagram}
Typically, the Ishikawa diagram \citep{tague2005quality} is applied as a cause analysis tool appropriate for describing the structure of a brainstorming session, in which a development team tries to identify possible reason causing a specific effect. Consequently, the Ishikawa chart is also called a cause/effect diagram. As a result of the brainstorming process, a fishbone diagram is constructed as an arrow with an arc directing to the effect (i.e., a problem statement). Then, the possible causes of the problem need to be identified that are presented as branches originating from the main arrow.

The diagram has also been applied in ARM visualization. For instance, the authors \cite{tsurinov2021farm} have established that ARM algorithms produce a large number of mined association rules in unstructured form. This means that there is no information about which features are more relevant for a user. In this sense, they proposed the Fishbone ARM (FARM) that is able to introduce a hierarchical structure for rules. The structure enables that the priority of features becomes clearly visible. 

The fishbone structure presents a basis for visualization with FARM. In this structure, features, inserted as ribs in a symbolic fishbone, are ordered such that the conviction metric values grow from the rear toward the head. Thus, the complexity of the structure increases by adding additional attributes. On the other hand, the statistical significance of the results also needs to be increased. In line with this, cross-validation is employed for evaluating the significance that splits the result dataset into two different portions (i.e., test and validation), and then re-sampled during more iterations.

\subsection{Molecular representation}
A molecule is a group of two or more atoms connected together with chemical bounds (e.g., covalent, ionic) \citep{ebbing2016general}.  Therefore, a molecule representation refers to a connected graph with nodes denoting atoms and edges denoting the chemical bounds between them. The representation  inspired \cite{said2013visualisation} into developing a new ARM visualization method that is devoted for visualizing items arising in the antecedent and consequent of the selected association rule. Thus, two characteristics need to be determined: (1) the contribution of each item to the rule, and (2) the correlation between each pair of antecedents and each pair of consequents from an archive of association rules. The association rules are explored before visualization according to one of the interestingness measures selected by the user, e.g., support, confidence, and lift.

The contribution of an item in the selected association rule $R=X\implies Y$ is calculated with measuring the Information Gain (IG) defined by~\citet{freitas1998on}:
\begin{equation}
    IG(A_i)=Info(R)-Info(R|A_i),
\end{equation}
where
\begin{equation}
    \begin{aligned}
        Info(R) &= -\sum_{j=1}^n{P(R_j)\log P(R_j)},~\text{and}\\
        Info(R|A_i) &= \sum_{k=1}^m{P(A{i,k})}\left(-\sum_{j=1}^n{P(R_j|A_{i,k})\log P(R_j|A_{i,k})}\right).
    \end{aligned}
\end{equation}
Thus, it holds that attributes with higher values of $IG$ are good predictors of the selected rule. In contrast, if items with low or negative $IG$ values are encountered, the selected rules are estimated as irrelevant. On the other hand, the lift interestingness measure (Eq.~\ref{Eq:lift}) is applied for determining the correlations between pairs of items in the antecedent and consequent, respectively.

The visualization of molecular representation is typically realized using sphere 3D graphs (also powered by R), where spheres present items and edges of the different distances' connection between them. The calculated characteristics of items into the selected rule are captured in a sphere graph as follows: 
\begin{itemize}
    \item the size of the sphere is proportional to the value of $IG$,
    \item the positive value of $IG$ is a plot in a sphere of one color (e.g., blue), while the negative one in a sphere of another color (e.g., white),
    \item the distance between two spheres is proportional to the measure lift.
\end{itemize}
However, authors \cite{said2013visualisation} simplified the visualization of association rules based on a molecular representation by developing a tool for VISual mining and Interactive User-Centred Exploration of Association Rules (IUCEARVis). 

In summary, the main weakness of the molecular structure is that it shows the importance of items to rules, and cannot show the distribution of association rules.

\subsection{Concept lattice}
A concept lattice is a tool for extracting specific information from massive data. It is obtained after a concept analysis that belongs to the domain of applied mathematics \citep{truong2010structure}. The results of the concept analysis are aggregated in a data structure that is, typically, presented in a Hasse graph. The Hasse graph consists of concepts representing as nodes in a 2-dimensional lattice, and edges expressing the generalization and instantiation of relationships between the concepts \citep{shen2020research}.

Formally, the concept lattice is defined as a triple $L=\langle O,A,B\rangle$, where $O$ denotes a set of objects, $A$ a set of attributes, and $B$ is a binary relationship matrix $B\subseteq O\times A$ denoting that an object $o\in O$ and attribute $a\in A$ are in a relationship, if $(i,a)\in B$. Thus, a node in the concept lattice is defined as a pair $\langle A,B\rangle$, where the former member is also called an extension $A\in O$ (i.e., a collection of objects), and the latter a connotation (i.e., collection of attributes). Indeed, a combination of objects and attributes is needed for a more comprehensive analysis of the association rules. 

The task of the ARM visual algorithms based on the context is to display association rules extracted from concept lattice. Thus, the central area of the visualization interface consists of a 2-dimensional lattice, within which the concepts are positioned as points according to their values of support and confidence. Two lines are attached below and above the lattice: The former represents the objects which have arisen in the antecedent, while the latter the same in the consequent of the potential association rule. Indeed, if there is a relationship between particular object and attribute in the relationship matrix $(i,a)\in B$, the object is connected with the node (concept) using an edge. 

The advantages of the ARM visualization based on a concept lattice can be summarized as follows: (1) a deeper understanding of association rules at the conceptual level, and (2) analyzing the relationships between concepts more comprehensively. However, the main weakness of the visualization is that this is only appropriate for visualizing the binary values of objects. In order to overcome the problem, \citet{yang2005pruning} proposed generalized association rules capable of visualizing the frequent rules in an itemset lattice that presents one item in parallel coordinates. In this way, many-to-many rules can be visualized on the one hand, and the large number of rules as selected by the user can be displayed on the other. Obviously, the advantage of the ARM visualization methods is that the user can limit the number of association rules for visualization interactively by specifying the parameters $S_{\min}$ and $C_{\min}$.

\subsection{Metro maps}
The concept of information maps enables analysis of data having a "geographical look" \citep{shahaf2012trains,shahaf2015metro}. The look can also be prescribed to mined association rules. Therefore, the idea to visualize these in the form  of metro maps has become appreciated \citep{fister2022information}. This means, similar as the metro map can help a user to orientate him/herself in the environment, the information map can help them to understand the information hidden in the mined association rules. Thereby, the metro map is divided into more metro lines, consisting of various metro stops. In the information sense, each metro stop represents an attribute, while the metro lines a linear sequence of the attributes (also different association rules). Mutual connections between the metro lines reveal how an attribute in one association rule affects an attribute in the other, and vice versa. Finally, understanding the linear sequences of attributes and connections between them can even tell stories about the specific information domain.

The metro map is defined mathematically as $\mathcal{M}=(G,\Pi$, where $G=(A,E)$ denotes an attribute graph of vertices $A=\{A_1,\ldots,A_M\}$, representing attributes and edges $E=\{r_i,\ldots,r_n\}$ representing simple rules (i.e., rules with one antecedent attribute and one consequent attribute), together with incident function $\psi_G$ that associates an ordered pair $\psi_G=(X,Y)$ denoting the implication $X\implies Y$, and $\Pi$ is a set of metro lines $\pi\in\Pi$ \citep{fister2022information}. The evolutionary algorithm was applied in~\cite{fister2022information} for constructing a metro map that must obey the following four objectives: (1) maximum path length $\tau$, (2) maximum map size $K$, (3) high coverage, and (4) high structure quality. 

Indeed, the maximum path length refers to the maximum number of metro stops (i.e., attributes) in a linear sequence. The maximum map size limits the number of metro lines. The coverage is proportional to the lift interestingness measure, where we were interested in rules with a lift value $>1$, determining the degree to which the probability of occurrence of the antecedent, and this of the consequent are dependent on one another. The structure quality ensures that the linear sequences of the metro stops are coherent in all metro lines.

An example of a metro map obtained by mining the Mushroom dataset, that was constructed using the parameters $\tau=6$ and $|\mathcal{K}|=4$, is illustrated in Fig.~\ref{fig:MM}.
\begin{figure}[htb]
\centering
\begin{minipage}{.6\textwidth}
  \includegraphics[width=.96\linewidth]{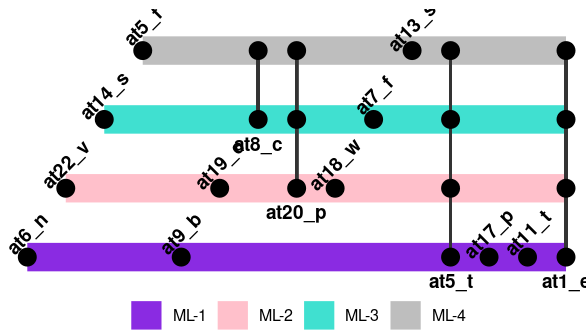}
  \caption{Metro map plot powered by R.}
  \label{fig:MM}
\end{minipage}
\begin{minipage}{.38\textwidth}
 \scriptsize
\begin{tabular}{ |c|l| }
\hline
Tag & Attribute \\	\hline
at1\_e & class\_edible \\
at1\_p & class\_poisonous \\
at5\_f & bruises?\_no \\
at5\_t & bruises?\_bruises \\
at6\_n & odor\_none \\
at7\_f & gill-attachment\_free \\
at8\_c & gill-spacing\_close \\
at9\_b & gill-size\_broad \\
at11\_t & stalk-shape\_tapering \\
at13\_s & stalk-surface-above-ring\_smooth \\
at14\_s & stalk-surface-below-ring\_smooth \\
at17\_p & veil-type\_partial \\
at18\_w & veil-color\_white \\
at19\_o & ring-number\_one \\
at20\_p & ring-type\_pendant \\
at22\_v & population\_several \\
\hline
\end{tabular}
\end{minipage}%
\end{figure}
Let us notice that the figure is divided into two parts, i.e., a diagram and a table. The diagram presents the visualized metro map, while the table the meaning of the metro stops (attributes).

\subsection{Sankey diagram}
Similar to the metro map, the Sankey diagram is also focused on "geographical data". Additionally, the kind of visualization enables visualization of hierarchical multivariate data. It is represented as a graph consisting of nodes representing attributes and edges representing connectivity by flows across time. In this diagram, the quality of each connection is distinguished by its weight that is proportional to some of the interestingness measures.

Mathematically, the Sankey diagram is defined as a directed graph $G=\langle K,R\rangle$, where $K$ denotes the maximum path length and $R$ is a set of similar rules~\cite{fister2022association}. The rules in this diagram are presented by the antecedent $X=\{A_{x_{1}}=a_{x_{1}}\wedge,\ldots,\wedge A_{x_{k}}=a_{x_{k}}\}$, representing a set of source nodes, consequent $Y=\{A_y=a_y\}$, representing a set of sink nodes, and interestingness measure $Z=\{\mathit{supp,cons,lift}\}$, reflecting the quality of a particular connection. The quality can also be expressed with a linear combination of the measures. The similarity between two rules $r_i$ and $r_j$ is defined as:
\begin{equation}
    \mathit{sim}(r_j,r_j)=\frac{|\mathit{Ante}(r_i)\cap \mathit{Ante}(r_j)|+|\mathit{Cons}(r_i)\cap \mathit{Cons}(r_j)|}{|\mathit{Ante}(r_i)\cup \mathit{Ante}(r_j)|+|\mathit{Cons}(r_i)\cup \mathit{Cons}(r_j)|},
\end{equation}
where $\mathit{Ante}(.)$ denotes a set of antecedent attributes, and $\mathit{Cons}(.)$ a set of consequent ones. However, the $\mathit{sim}(r_i,r_j)\in [0,1]$, where the value 0 means that the rules are not similar, and 1 that the rules are absolutely similar. The similarities are then combined into an adjacency matrix $\mathit{Adj}$, defined as follows:
\begin{equation}
    \text{Adj}=\left[ \begin{matrix}
    a_{1,1} & \ldots & a_{1,M} \\
     & \ldots &  \\
    a_{\tilde{M},1} & \ldots & a_{M,M}. \\
    \end{matrix} \right],
\end{equation}
The problem of searching for the most similar set of association rules $R$ is defined as a Knapsack 0/1 problem \citep{kellerer2010knapsack}. 

The construction of the Sankey diagram visualization is divided into two steps: (1) searching for a set of the most similar association rules, and (2) visualization using Sankey diagrams. In~\citet{fister2022association}, the authors proposed a DE meta-heuristic algorithm using the Knapsack 0/1 deterministic algorithm for determining the set of the most similar rules, while the R programming language for statistical computing was applied to solve the second step.

The example of Sankey diagrams is illustrated in Figs.~\ref{fig:river_plot1}-\ref{fig:river_plot2} 
\begin{figure*}[!ht]
\begin{subfigure}[t]{.48\linewidth}
  \includegraphics[width=\textwidth]{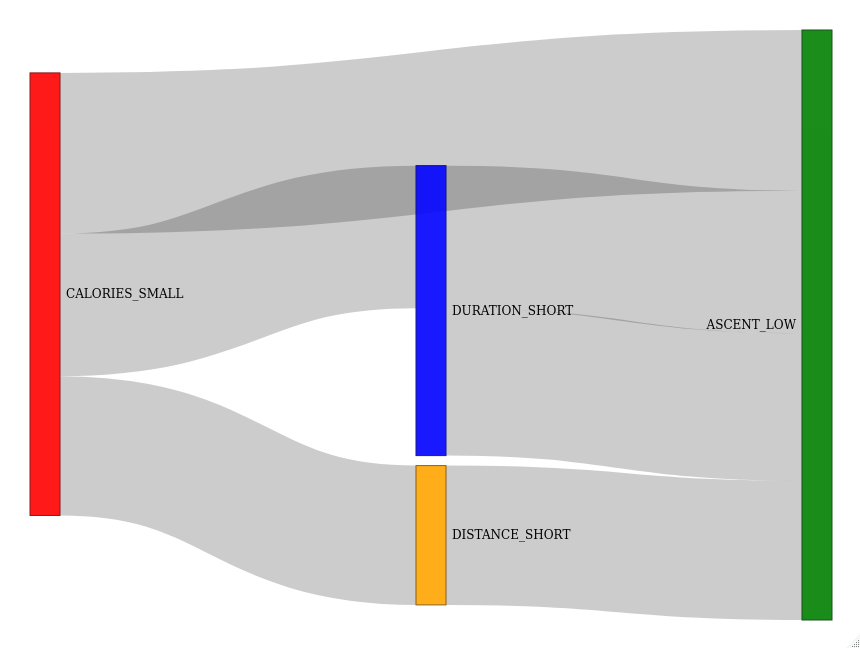}
  \caption{Sankey diagram for time period 1.}
  \label{fig:river_plot1}
\end{subfigure}
\hfill
\begin{subfigure}[t]{.48\linewidth}
  \includegraphics[width=\textwidth]{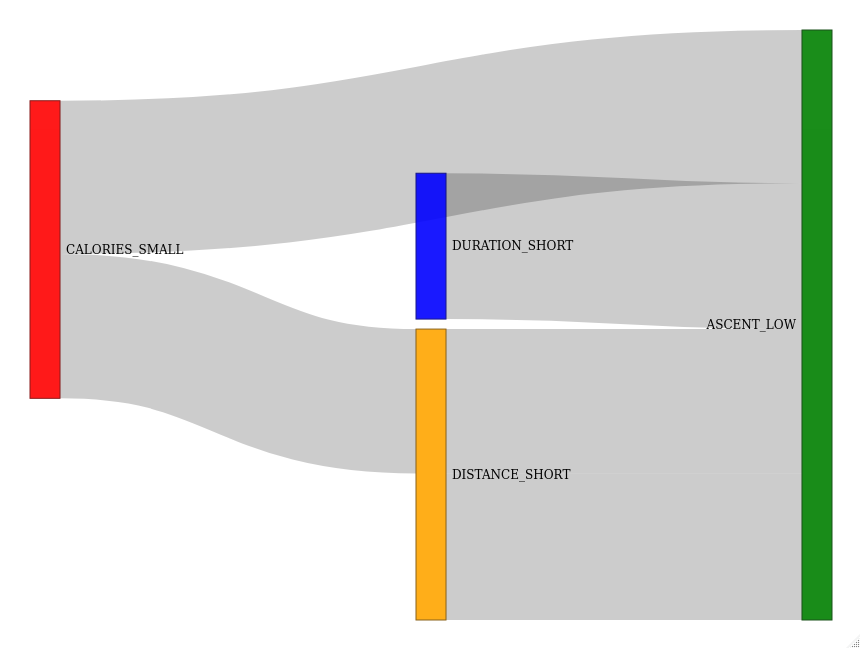}
  \caption{Sankey diagram for time period 2.}
  \label{fig:river_plot2}
\end{subfigure}
\caption{Sankey diagrams for time periods 1 and 2 powered by R.}
\end{figure*} 
that refer to mining the sport training database obtained in more seasons (i.e., years). This database consists of training load indicators measured during an implementation of a sport training session. The visualization is divided into two parts: The first part (Fig.~\ref{fig:river_plot1}) presents the results of the ARM visualization on sport training data captured during one season, while the second (Fig.~\ref{fig:river_plot2}) highlights the data obtained during the next season. 

In this way, two historical insights are served to a sport trainer: (1) In what proportion do the training load indicators contribute to the whole? and (2) What changes can be observed in the sense of training load indicators by athletes who have already had the main portion of training sessions during the previous seasons?

Interestingly, \cite{hlosta2013approach} proposed a visualization of evolving association rules using graphs, where the nodes of the graphs represent items and edges specific association rules. Thus, the graph-based diagram shows how evolving models mined using the ARM algorithms and stored into a transaction database can be filtered and visualized.

\subsection{Ribbon plot}
Ribbon plots are appropriate for visualizing data without self-intersections, where linearized simplification of events exposes the significant ones. Although the plot is ideal for analyzing linearized sequences, it can be applied successfully for visualizing the best association rule in NARM, where the proper boundaries need to be discovered between the numerical attributes. Thus, the attribute with the best support is compared with the other attributes in the association rule according to support and confidence. The attributes are ordered into linear sequence according to the closeness of the first attribute regarding the others.

The inspiration behind the visualization is presented by the Tour De France (TDF), i.e., the most famous cycling race in the world. Similar as in the TDF, where the best hill climbers have more chance to win the race, the attribute with the higher support also has the most decisive role in a decision-making process. Indeed, virtual hill slopes are visualized as triangles situated on a plain, where the left leg denotes an ascent and the right leg a descent of the virtual hill in a linear sequence, starting from the left to the right side. In the paper of~\citet{fister2020visualization}, the ascent of the virtual hill is proportional to the attribute's support, while the descent to the confidence of the simple association rule.

Mathematically, the best rule $X\Rightarrow Y$ consists of an antecedent $X=\{A_x=a_x\}$ and a consequent $$Y=\{A_{y_1}=a_{y_1},\ldots,A_{y_k}=a_{y_k}\}$$, where the $A_x$ denotes the best attribute according to the support, and simple association rules $A_x\Rightarrow A_{y_j}$ for $j=1,\ldots,k$ are ordered as:
\begin{equation}
    \mathit{conf}(A_x\Rightarrow A_{y_{\pi_1}})\geq \mathit{conf}(A_x\Rightarrow A_{y_{\pi_k}}),
\end{equation}
where $\pi_j$ is a permutation of the attributes belonging to the consequent. Moreover, the distances $\mathit{dist}_j$ between the virtual hills are also proportional to $\mathit{dist}_j\propto \mathit{conf}(A_x\Rightarrow A_{y_{\pi_j}})$. 

An example of a ribbon plot is illustrated in Fig.~\ref{fig:Ribbon} representing a visualization of the best association rule mined by the uARMSolver \citep{fister2020uarmsolver} (i.e., the framework for NARM using the nature-inspired algorithms). 

\begin{figure}[htb]
\centering
\begin{subfigure}[t]{.48\linewidth}
  \centering
  \includegraphics[width=1.\linewidth]{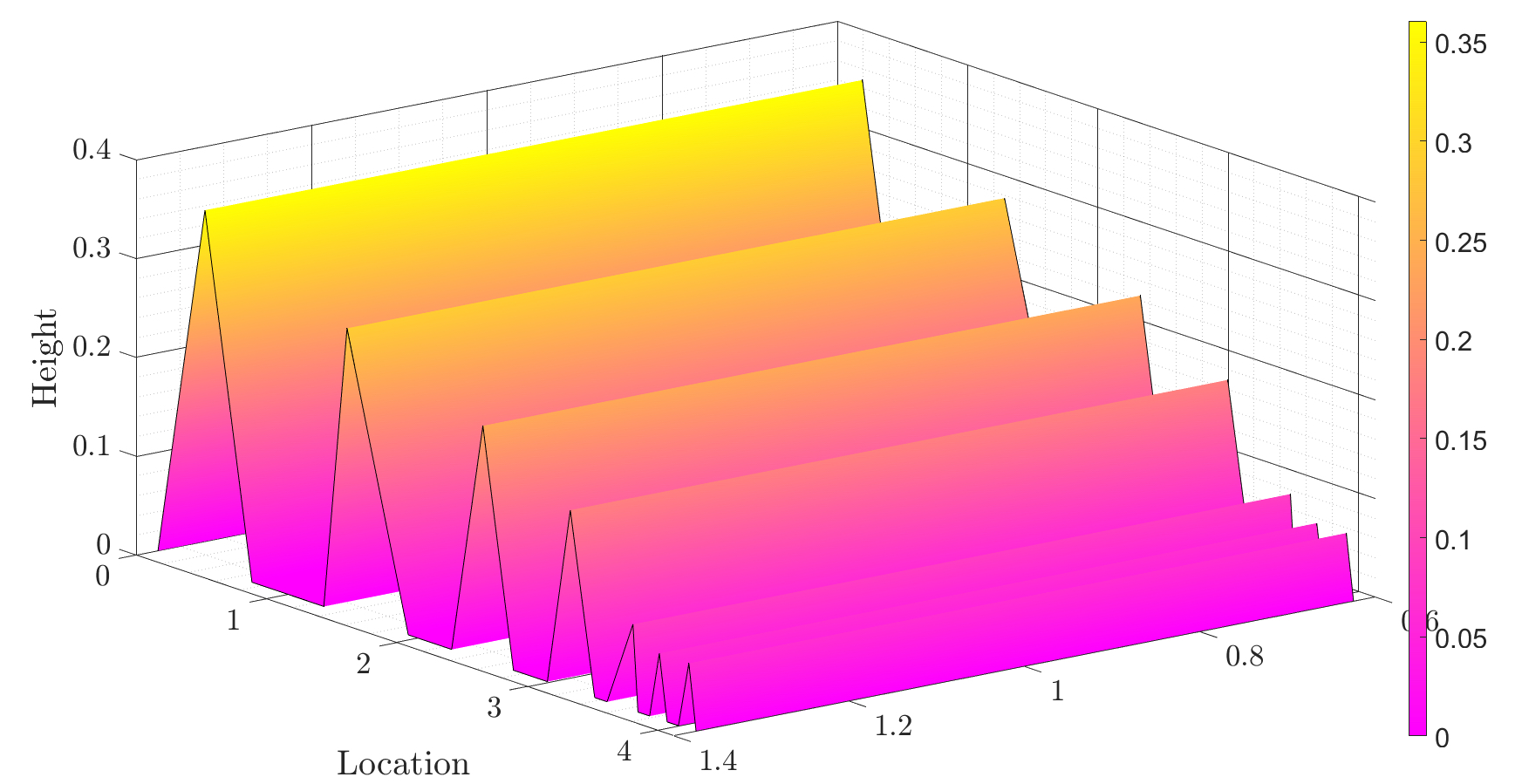}
  \caption{Ribbon plot powered by Matlab.}
  \label{fig:Ribbon}
\end{subfigure}
\begin{subfigure}[t]{.48\linewidth}
 \centering
  \includegraphics[width=1.\textwidth]{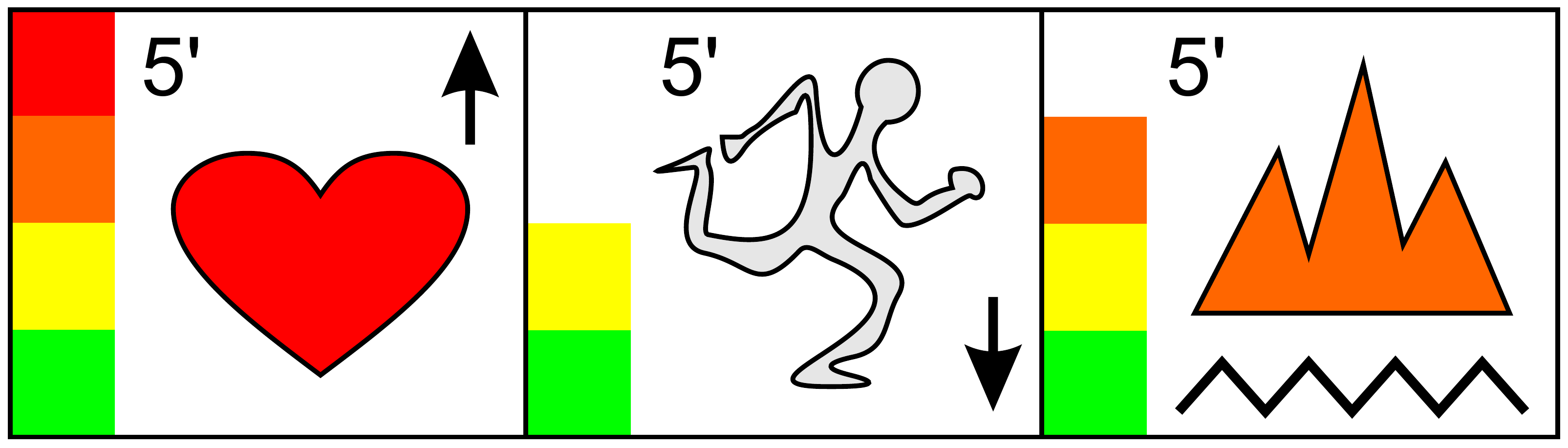}
  \caption{Glyph-based chart.}
  \label{fig:glyph}
\end{subfigure}
\caption{Ribbon plot and Glyph-based chart.}
\end{figure}
The framework was applied for mining a database consisting of transactions obtained by cycling training sessions. Thus, the best transaction is composed from seven attributes $A_1,\ldots,A_{k+1}$ ordered into the association rule:
\begin{equation*}
    A_x\Rightarrow A_{y_{\pi_1}}\wedge \ldots\wedge A_{y_{\pi_k}}.
\end{equation*}
Seven virtual hills can be observed as can be seen from the figure. While the first three virtual hills are of comparable height to the first one, the remainder of the hills are of lower height, and, thus, reflect the lower inter-dependence. 

\subsection{Glyph-based plots}
Glyph-based plots are suitable for visualizing multivariate data with more than two attribute dimensions, where different data variables are presented by a set of visual channels (i.e., shape, size, color, orientation, curvature, etc.) \citep{borgo2013glyph}. Indeed, glyphs are devoted for depicting attributes of data that, typically, appear in collections of visualized objects. They are founded on the basics of a semiotic theory that is, in fact, the science of signs \citep{lagopoulos2020theory}. According to this theory, signs have emerged in three forms: icons, indices, and symbols. Icons reflect a physical correlation to the sign. The index expresses a space and time correlation to the object. In other words, they have an indirect effect on the object. A meta-physic correlation (i.e., no real correlation) exists between the symbol and the sign.

An example of glyph-based visualization for ARM was performed by~\citet{hrovat2015interestingness} that analyzed the time series data gathered from a single athlete (i.e., a cyclist) during a large time period of training (i.e., the whole season). In this study, the sequential pattern mining algorithm \citep{agrawal1994fast} was exploited, where the sequential patterns were discovered by employing the novel trend interestingness measure for mining sequential patterns. Thus, a time-series sequences $ts=\langle,ts_1,\ldots, ts_m\rangle$ were discovered from a transaction database consisting of sport training performed by a single athlete.

Two trend interestingness measures are defined in the study as follows: (1) the duration trend $\overrightarrow{\mathit{dut}}(ts)$, and (2) the daily trend $\overrightarrow{\mathit{dat}}(ts)$. The former discovers trends within a trend database on a monthly, while the latter on a daily basis. The trend database is constructed from the original transaction database by dividing each training session into $m$-time series. Then, the permutation test is performed, after which those sequential patterns are selected with a minimum $p$-value. Obviously, the $p$-value is obtained as a result of the permutation test, and serves as a trend interestingness measure. 

Both trend interestingness measures are visualized using glyphs in order to depict how trends increase or decrease during a specific training period (Fig.~\ref{fig:glyph}). Thus, two glyph symbols are used by the visualization: (1) level, and (2) variable. The level's symbol depicts the trend interestingness measure using an optical channel (i.e., color), where the intensity training load indicators are presented in different colors, depending on low, moderate, intensity, and high intensity levels. The variable's symbol addresses the geometric channels, like: the cyclist's speed (as maximum, average or standard deviation), average heart rate (as minimum, maximum, average and standard deviation), and altitude (as standard deviation). These symbols are depicted using different shapes. 

\subsection{Other ideas in ARM visualization}
The characteristics of the remainder of the analyzed papers can be summarized in the present section as follows: The majority of the papers were published for various data mining conferences. As a result, these include ideas more on the conceptual level, and, therefore, the solutions that they reveal are not robust enough for using in the everyday real-world environment. On the other hand, these ideas are not included into some recognizable ARM visualization system. However, they could be interesting for the potential readers for sure.

The principles of ARM visualization methods, as found in the observed collections of analyzed papers, can be classified into the following two classes (Table~\ref{tab:2a}):
\begin{itemize}
    \item reducing a rule set,
    \item visual data mining.
\end{itemize}
Indeed, the first two principles of ARM visualization are used commonly in the ARM community: Thereby, the association rules are mined using some of the known mining algorithms. These algorithms produce a lot of association rules that need to be reduced (also rummaged) into an association rule set necessary for visualization. The second principle is more goal-oriented, and mines the association rules either in a visualization context, or tries to reduce their number by avoiding occlusions using optimization. In this way, the association rule set does not need to be reduced further before visualization. Interestingly, the first principle is characterized for papers which emerged at the beginning of the ARM visualization domain, while the second one is typical for papers of the new age, where the ARM exploration is part of the visualization process.

\begin{table}[htb]
\caption{Other ARM visualization methods.}
\label{tab:2a}
\small
\begin{tabular}{ c|l|l|r }
\hline
\multicolumn{3}{c}{Principles of ARM visualization} \\\cline{1-4}
\multicolumn{2}{l|}{1 Reducing rule set} & Attribute & Reference\\\hline 
1 & Relational SQL & Categorical & \cite{chakravarthy2003visualization} \\
2 & Conditional AR analysis & Categorical & \cite{yamada2015visualization} \\
3 & Rummaging model & Categorical & \cite{blanchard2003user} \\
4 & Rummaging model & Categorical & \cite{menin2021arviz} \\
5 & Correlation rule visualization & Categorical & \cite{zheng2017visualization} \\
6 & Weighted association rules & Categorical & \cite{saeed2011activity} \\
7 & Multiple antecedent & Text & \cite{wong1999visualizing} \\
8 & Weighted association rules & Text & \cite{kawahara1999performance} \\
9 & Hierarchical structure & Boolean & \cite{jiang2008finite} \\
\hline   
\multicolumn{3}{l}{2 Visual data mining} \\\hline
1 & Correlation visualization alg. & Categorical & \cite{xu2009visualization} \\
2 & Integrated framework & Categorical & \cite{couturier2007scalable} \\
3 & Occlusion reducing & Categorical & \cite{couturier2008optimizing} \\
4 & Contextual exploration & Categorical & \cite{yahia2004contextual} \\
5 & Generic association rule set & Categorical & \cite{yahia2004emulating} \\
6 & 3D visualization engine & Categorical & \cite{ounifi2016new} \\
7 & Rule-to-items mapping & Categorical & \cite{wang20173d} \\
\hline   
\end{tabular}
\end{table}

Obviously, the reducing can be performed on many ways. For instance, \cite{chakravarthy2003visualization} proposed a relational SQL query language, with which a user can select the suitable association rule set for visualization interactively from the collection of association rules stored in tables. \cite{yamada2015visualization} applied the conditional association rule analysis and the association rule analysis with user attributes for the comprehending questionnaire data. \cite{blanchard2003user} introduced the rummaging model for filtering association rules interactively, and included into an experimental prototype called ARVis. A similar model was recommended by \cite{menin2021arviz}, devoted to exploring the RDF data that employed the traditional methods for visualizing and was incorporated into the prototype ARViz. The gray correlation rule visualization algorithm was advised by \cite{zheng2017visualization} that is suitable for considering the influence of the association rules on the visualization. \cite{saeed2011activity} mined a collection of documents consisting of metadata with the Apriori algorithm, and selected an association rule set for visualization according to the calculated weights. 

On the other hand, \citet{wong1999visualizing} visualized an association rule set with multiple antecedents using a 3-dimensional graph, and applied their solution to a text mining study on a large corpora. \cite{kawahara1999performance} developed the web search engine for manipulating weighted association rules. Thus, the text mining algorithm derived appropriate keywords, while the ROC graph served for the visualization of association rules. Boolean association rules were visualized by~\citet{jiang2008finite} using the hierarchical structure for all of them and depicted in a Hasse diagram. 

Visual data mining can be performed in various ways, as found in our study: The correlation visualization algorithm was proposed for mining the alarm association rules by \cite{xu2009visualization}. \cite{couturier2007scalable} recommended the integrated framework for association rule extraction and visualization in one step, which integrated previous methods of association rule visualization. Occlusion optimization was proposed by \cite{couturier2008optimizing}.  
Contextual exploration of an association rule set was developed by \cite{yahia2004contextual} and \cite{yahia2004emulating}, where the additional knowledge needed for visualization was constructed using the fuzzy meta-rules. \cite{ounifi2016new} solved the problem of extraction and visualization by a 3-dimensional visualization engine, while \cite{wang20173d} introduced a 3-dimensional matrix-based visualization system, where the basic matrix-based approach was extended by rule-to-items mapping.

\subsection{Taxonomies of the ARM visualization}
The ARM visualization methods can be classified according to many aspects. These aspects depend on the various standpoints from which they are observed. Indeed, the following questions reflect those standpoints more precisely: 
\begin{itemize}
    \item How to visualize?
    \item Which visualization methods to use?
    \item Which characteristics of association rules are essential to visualize?
    \item What to visualize?
    \item Which type of attributes to display?
\end{itemize}
In the remainder of the section, these queries are described in detail.

\subsubsection{How to visualize?}
The aspect "How to visualize?" refers to the mode of how the exploration and visualization are performed. In line with this, four different methods are distinguished, as follows (Fig.~\ref{fig:tax_1}):
\begin{itemize}
\item reducing the item set,
\item visual data mining,
\item a concept lattice,
\item evolving association rules.
\end{itemize}
Reducing the item set means that the exploration of association rules is performed with traditional ARM methods (e.g., Apriori, Eclat, evolutionary algorithms), after which the visualization is performed using some traditional or new age visualization methods. 
\begin{figure}
  \includegraphics[width=.9\textwidth]{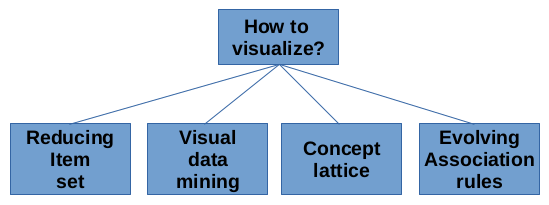}
  \caption{How to visualize?}
  \label{fig:tax_1}
\end{figure}
Visual data mining comprises those ARM visualization methods that perform the exploration and visualization phases in one step. These methods mine association rules more directionally, where mining can be performed from some concept, can use meta rules, or can be able to limit the number of occlusions. The concept lattice enables displaying the structure of the association rules (i.e., attributes) beside the single rules. However, this visualization method is reserved for displaying the binary association rules only. The evolving association rules are appropriate for visualizing either warehouse data cubes stored in a multidimensional data model, or data suitable for displaying by Sankey diagrams. 

\subsubsection{Which visualization methods to use?}
This aspect is focused on the question, which visualization method to use? In line with this, we can consider that the methods are divided into traditional and new age visualization methods. The former consists of charts, like scatter plot, group-based, matrix-based and mosaic plots, and their variants, like two-key, grouped-matrix and double Decker plots (see Table~\ref{fig:taxonomy} under the column "Method"). The new age visualization methods are comprised of an Ishikawa diagram, molecular representation, a concept lattice, metro maps, Sankey diagrams, ribbon plots, and glyph-based charts.

\begin{scriptsize}
  \begingroup 
    \renewcommand*{\arraystretch}{1.5}%
    \definecolor{tabred}{RGB}{230,36,0}%
    \definecolor{tabgreen}{RGB}{0,116,21}%
    \definecolor{taborange}{RGB}{255,124,0}%
    \definecolor{tabbrown}{RGB}{171,70,0}%
    \definecolor{tabyellow}{RGB}{255,253,169}%
    \definecolor{tabviolet}{rgb}{0.6, 0.4, 0.8}
    \newcommand*{\redtriangle}{\textcolor{tabred}{\ding{115}}}%
    \newcommand*{\greenbullet}{\textcolor{tabgreen}{\ding{108}}}%
    \newcommand*{\greenone}{\textcolor{tabgreen}{\ding{182}}}%
    \newcommand*{\greencouple}{\textcolor{tabgreen}{\ding{183}}}%
    \newcommand*{\greenfew}{\textcolor{tabgreen}{\ding{184}}}%
    \newcommand*{\greenfewless}{\textcolor{tabgreen}{\ding{174}}}%
    \newcommand*{\greenseveral}{\textcolor{tabgreen}{\ding{185}}}%
    \newcommand*{\greenmany}{\textcolor{tabgreen}{\ding{186}}}%
    \newcommand*{\orangecirc}{\textcolor{taborange}{\ding{109}}}%
    \newcommand*{\violetsquare}{\textcolor{tabviolet}{\ding{110}}}%
    \newcommand*{\headformat}[1]{{\small#1}}%
    \newcommand*{\vcorr}{%
      \vadjust{\vspace{-\dp\csname @arstrutbox\endcsname}}%
      \global\let\vcorr\relax
    }%
    \newcommand*{\HeadAux}[1]{%
      \multicolumn{1}{@{}r@{}}{%
        \vcorr
        \sbox0{\headformat{\strut #1}}%
        \sbox2{\headformat{Complex Data Movement}}%
        \sbox4{\kern\tabcolsep\redtriangle\kern\tabcolsep}%
        \sbox6{\rotatebox{45}{\rule{0pt}{\dimexpr\ht0+\dp0\relax}}}%
        \sbox0{\raisebox{.5\dimexpr\dp0-\ht0\relax}[0pt][0pt]{\unhcopy0}}%
        \kern.75\wd4 %
        \rlap{%
          \raisebox{.25\wd4}{\rotatebox{45}{\unhcopy0}}%
        }%
        \kern.25\wd4 %
        \ifx\HeadLine Y%
          \dimen0=\dimexpr\wd2+.5\wd4\relax
          \rlap{\rotatebox{45}{\hbox{\vrule width\dimen0 height .4pt}}}%
        \fi
      }%
    }%
    \newcommand*{\head}[1]{\HeadAux{\global\let\HeadLine=Y#1}}%
    \newcommand*{\headNoLine}[1]{\HeadAux{\global\let\HeadLine=N#1}}%
    \noindent
    \begin{table}[htb]
    \begin{scriptsize}
    \begin{tabular}{%
      >{\bfseries}l|>{\quad}
      *{3}{c|}>{\quad}c|
      *{5}{c|}>{\quad}c|
      *{3}{c|}c%
    }%
      &
      \head{Nr. of interesting meas.} &
      \head{Rule set size} &
      \head{Interactive} &
      &
      \head{Interesting measure} &
      \head{Rule length} &
      \head{Items} &
      \head{RHS+LHS} &
      \head{Time-series} &
      &
      \head{Categorical} &
      \head{Numerical} &
      \head{Binary} &
      \\
      \hline
      \sbox0{S}%
      \rule{0pt}{\dimexpr\ht0 + 2ex\relax}%
      Scatter plot & \textcolor{tabgreen}{\bfseries 3} & \greenseveral & \greenbullet && \redtriangle & &  & & && \violetsquare &  & \\
      Two key plot & \textcolor{tabgreen}{\bfseries 2+} & \greenseveral & \greenbullet & && \redtriangle & & &  && \violetsquare &  & \\
      Graph-based & \textcolor{tabgreen}{\bfseries 2} & \greenfew & & &&  & & \redtriangle & && \violetsquare &  & \\
      Matrix-based & \textcolor{tabgreen}{\bfseries 1} & \greenfewless & \greenbullet &&  & & & \redtriangle & && \violetsquare &  & \\
      Grouped matrix & \textcolor{tabgreen}{\bfseries 2} & \greenmany & \greenbullet &&  & & & \redtriangle & && \violetsquare &  & \\
      Mosaic plot & \textcolor{tabgreen}{\bfseries 2} & \greenone & \greenbullet &&  & & & \redtriangle & && \violetsquare &  & \\
      Double Decker & \textcolor{tabgreen}{\bfseries 2} & \greenone & \greenbullet &&  & & & \redtriangle & && \violetsquare &  & \\
      \hline
      Ishikawa diagram & \textcolor{tabgreen}{\bfseries 1} & \greencouple & &&  & & \redtriangle &  & && \violetsquare & \violetsquare & \\      
      Molecular representation & \textcolor{tabgreen}{\bfseries 3} & \greenone & \greenbullet && & & \redtriangle &  & && \violetsquare &  & \\      
      Concept lattice & \textcolor{tabgreen}{\bfseries 1} & \greencouple & && & & \redtriangle & \redtriangle & &&  &  & \violetsquare\\  
      Metro map & \textcolor{tabgreen}{\bfseries 1} & \greencouple & && & & \redtriangle &  \redtriangle & && \violetsquare & \violetsquare & \\      
      Sankey map & \textcolor{tabgreen}{\bfseries 2} & \greencouple & && & &  & \redtriangle & &&  & \violetsquare & \\      
      Ribbon plot & \textcolor{tabgreen}{\bfseries 2} & \greenone & && & &  &  \redtriangle & && \violetsquare & \violetsquare & \\      
      Glyph-based chart & \textcolor{tabgreen}{\bfseries 1} & \greenone & &&  & & & & \redtriangle && \violetsquare & \violetsquare & \\      \hline
      \multicolumn{1}{c|}{\bfseries Method} &
      \multicolumn{4}{c|}{\bfseries Characteristics} &
      \multicolumn{5}{c|}{\bfseries Focus} &
      \multicolumn{3}{c}{\bfseries Attribute}
      \\
    \end{tabular}%
    \caption{Taxonomy of the ARM visualization methods.}
    \label{fig:taxonomy}
    \end{scriptsize}
    \end{table}
    \kern19.5mm 
  \endgroup

\end{scriptsize}

\subsubsection{Which characteristics of association rules are essential to visualize?}
The characteristics of the ARM visualization methods refer to: (1) the number of displayed interestingness measures, (2) the rule set size, and (3) the interactivity tools. The number of displayed interestingness rules determines, how many of the interestingness measures are included into the representation for user. For instance, the scatter plot is able to display three interestingness measures, while the two-key plot actually only two, but the third measure is presented indirectly by a color. In general, the number of measures by various visualization methods are typically in the range $[1,3]$. The rule set size determines the number of association rules to be displayed by the definite visualization method. This number is denoted in Table~\ref{fig:taxonomy} in the column "Rule set size" in circles with numbers within them. The numbers present the powers of base 10. This means that the grouped matrix can display $10^5$ association rules. The column "Interactive" shows if specific visualization method supports interactive tools (e.g., hover, zoom, pan, drill down, etc.) or not. Interestingly, although the new age visualization methods do not support interactive tools in general, they allow tuning of parameter settings that enable users some kind of interactivity. 

\subsubsection{What to visualize?}
The aspect, answering to the question "What to visualize?", deals with the focus, which an ARM visualization is presenting. Actually, the ARM visualization can be focused on illustrating: (1) number of interestingness measures, (2) rule length, (3) items, (4) RHS and LHS, and (5) time series data. The first focus is devoted to displaying the number if interestingness measures. The rule length refers to the number of attributes in the visualized association rules. The item focuses on depicting the attributes of the association rules, while the RHS+LHS focus is concentrated on the structure of the more important rules. Finally, the last focus considers the time series data. 

Interestingly, the concept lattice and metro maps even cover two focuses of displaying association rules, i.e., items (i.e., attributes) and their structure. On the other hand, the glyph-based visualization is dedicated for presenting the time series data.

\subsubsection{Which type of attributes to display?}
The aspect "Which type of attributes to display?" is focused on visualization based to distinguish the attribute types. In the ARM exploration/visualization, three attribute types can be identified as follows: (1) categorical, (2) numerical, and (3) binary. Interestingly, the majority of the traditional visualization methods are suitable for displaying the categorical type of attributes. Usually, displaying attributes of the numerical type is performed by these visualization methods by discretizing the numerical attributes into discrete classes. Obviously, the new age visualization methods are capable of working with the numerical and binary attributes directly as well.

\section{ARM visualization systems}\label{sec:6}
The section aims to compile a list of specialized ARM visualization systems and software packages for any of the ARM visualization methods. Obviously, this does not present the other visualization libraries, from which we can develop some methods (e.g., \textbf{matplotlib} in Python, or \textbf{ggplot2} in R). The study focused on presenting only the collection of graphics system that are used more nowadays in the ARM community. The collection of systems is illustrated in Table~\ref{tab:3}.

\begin{table}[htb]
\begin{center}
\caption{List of the ARM graphics systems.}
\label{tab:3}
\begin{tabularx}{\textwidth}[t]{XX}
\arrayrulecolor{black}\hline
\textbf{\textcolor{novabarva}{R packages}} & \\
\arrayrulecolor{novabarva}\hline
arulesViz \url{https://cran.r-project.org/web/packages/arulesViz/index.html} & 
\begin{minipage}[t]{\linewidth}%
\begin{itemize}
\item[1.1] probably the only state-of-the-art tool that supports many visualization methods up to this date
\item[1.2] includes also interactive tools
\end{itemize} 
\end{minipage}\\

\arrayrulecolor{black}\hline

\hline

\arrayrulecolor{green}\hline
\textbf{\textcolor{novabarva}{Python packages}} \\
\arrayrulecolor{black}\hline

pycaret \url{https://github.com/pycaret/pycaret} &
\begin{minipage}[t]{\linewidth}%
\begin{itemize}
\item[2.1] basically low-code machine learning library in Python
\item[2.2] association rule mining is a part of this library
\item[2.3] library supports 2D and 3D plots of association rules
\end{itemize}
\end{minipage}\\

\arrayrulecolor{novabarva}\hline

NiaARM (\url{https://github.com/firefly-cpp/NiaARM})&
\begin{minipage}[t]{\linewidth}%
\begin{itemize}
\item[3.1] minor module devoted for visualization
\item[3.2] for now supports only ribbon plots
\end{itemize}
\end{minipage}\\

\arrayrulecolor{novabarva}\hline

PyARMViz \url{https://github.com/Mazeofthemind/PyARMViz} &
\begin{minipage}[t]{\linewidth}%
\begin{itemize}
\item[4.1] Python Association Rule Visualization Library that is loosely based on ArulesViz
\item[4.2] Development probably stalled (no commits in the last 2.5 years)
\end{itemize}
\end{minipage}\\

\arrayrulecolor{black}\hline

\hline
\arrayrulecolor{black}\hline
\multicolumn{2}{l}{%
\textbf{\textcolor{novabarva}{C++ packages}}} \\
\arrayrulecolor{black}\hline

uARMSolver~\url{https://github.com/firefly-cpp/uARMSolver} &
\begin{minipage}[t]{\linewidth}%
\begin{itemize}
\item[5.1] small part of this package is devoted to the visualization
\item[5.2] provides the coordinates for metro plots which can be later visualized using metro map algorithms
\end{itemize} 
\end{minipage}
\end{tabularx}
\end{center}
\end{table}

As can be seen from the table, the \textbf{arulesViz} graphics system is the most complete, due to covering the majority of the visualization methods dealt with in this review paper. This is an extensive toolbox in the R-extension package \citep{hahsler2011arules}, and works in two phases: (1) exploration using known ARM methods to which tools for reducing the huge number of association rules are applied (e.g., filtering, zooming and rearranging), and (2) visualization of results. The current version of the software supports the following visualization methods (i.e., graphics): scatter plots, network plots, matrix-based, graph-based, mosaic plots and parallel coordinate plots.  

The other libraries are just a smaller drop in the ocean and, typically, they solve only limited ARM exploration/visualization approaches. For example, while the \textbf{NiaARM} is focused at this moment on only one visualization method (i.e., ribbon plot), the \textbf{PyARMviz} graphics system tends to be what is arulesViz for R, but in Python. Unfortunately, the development of this graphics software has probably stalled since the last commit was done almost three years ago. On the other hand, the development of the NiaARM is not finished yet, due to the unfinished inclusion of the new ideas in ARM visualization (e.g., metro maps, Sankey diagram, etc.) that should shortly widen the usability of the graphics system.

\section{Challenges and open problems}\label{sec:7}
After deep analysis of the ARM visualization methods, we can conclude that a universal method for covering all the ARM visualization problems does not exist. As a result, the arulesViz software package offers a spectrum of solutions useful for visualization with traditional ARM visualization methods. In this package, the scatter plot is applied as an entry point for an analysis of how to distinguish the similarity of association rules according to interestingness measures, like support and confidence. Then, the matrix-based visualization can be applied, capable of organizing association rules into a matrix, where the antecedent and consequent items can be distinguished. Finally, the graph-based methods are recommended by authors, in order to get the user the broadest view of the relationships between individual items reflecting, their memberships in different association rules. 

In summary, the problems caused by using the traditional ARM visualization methods can be aggregated as follows \citep{shen2020research}:
\begin{itemize}
    \item the domain knowledge is not displayed sufficiently, i.e., the rules are displayed from a single point of view,
    \item the visualization of background knowledge is not enough for sharing, i.e., the role and relationship of global information is lost in the context of the background knowledge,
    \item the use and exploration of potential knowledge hidden in non-connected attributes are reduced.
\end{itemize}
However, the new age ARM visualization methods tries to reveal the aforementioned problems. Moreover, some of these methods are even able to tell stories in mined data (e.g., metro maps), while the others are able to analyze the information from the history point of view (e.g., Sankey diagrams).

Although searching for a new age ARM visualization methods almost stopped after the rapid development of the traditional ARM visualization methods in the past, in our opinion, the future of the ARM visualization remains in the development of the new age ARM visualization methods. These methods might consolidate displaying items as well as the structure of the association rules. Additionally, these need to be independent of the attribute types.

The main advantage of the ARM visualization undoubtedly presents the interactivity of the ARM visualization methods. Interactive visualization improves the user's experience and interpretation of the results. Although several popular implementations of the traditional ARM visualization methods (e.g., arulesViz R-package by \cite{hasler2017visualizing}, and InterVisAR by \citep{intervisar}) already offer some interactive tools (e.g., hover, zoom, pan, drill down, inspect, brush), these tools are usually missing in the observed new age ARM visualization methods.

\section{Conclusions}\label{sec:8}
Data mining methods today suffer from a lot of comprehension of the mass results they produce. In line with this, a new domain of AI, the so-called XAI, has emerged that searches for methods which will be suitable to present these results clearly to the user. The visualization methods are one of the useful tools for helping users understand the results of different data mining methods better.

The present study has revised the most important visualization methods associated with ARM. Consequently, the most important ARM visualization methods, published in research papers, have been identified, analyzed, and classified. The ARM visualization methods are divided into traditional and new age methods. Moreover, they have been classified according to the characteristics of the displayed association rules, the focus of visualization, and the types of attributes.

The potential reader of this work will be able to get deeper overview of the ARM exploration/visualization process. Furthermore, it encourages readers to open new avenues of potential research. According to the research paper review, there is a huge opportunity to use the knowledge, especially in biological/medical sciences.

\section*{Acknowledgements}
This research has been supported partially by the project PID2020-115454GB-C21 of the Spanish Ministry of Science and Innovation (MICINN).). The authors acknowledge the financial support from the Slovenian Research Agency (Research Core Funding No. P2-0057 \& P2-0042).

\bibliography{sample}

\end{document}